\documentclass[twocolumn,floats,showpacs,prd,amssymb,floatfix,nofootinbib,balancelastpage,superscriptaddress,amsmath]{revtex4}
\usepackage{psfig}

\newcommand{\sss}{\scriptscriptstyle}

                              \newlength{\strikewidth}
                              \newlength{\strikelength}
\begin{document}

\title{The Shadow of Dark Matter}

\author{Stefano Profumo}
\email{profumo@caltech.edu}
\affiliation{California Institute of Technology, MS-106-38, Pasadena, CA 91125, USA}
\author{Kris Sigurdson}
\thanks{Hubble Fellow}
\email{krs@ias.edu}
\affiliation{School of Natural Sciences, Institute for Advanced Study, Princeton, NJ 08540, USA}
\affiliation{Department of Physics and Astronomy, University of British Columbia, Vancouver, BC V6T 1Z1, Canada}

\begin{abstract}
\noindent We carry out a model independent study of resonant photon scattering off dark matter (DM) particles. The DM particle $\chi_1$ can feature an electric or magnetic transition dipole moment which couples it with photons and a heavier neutral particle $\chi_2$. Resonant photon scattering then takes place at a special energy $E_\gamma^{\rm res}$ set by the masses of $\chi_1$ and $\chi_2$, with  the width of the resonance set by the size of the transition dipole moment. We compute the constraints on the parameter space of the model from stellar energy losses, data from SN 1987A, the Lyman-$\alpha$ forest, Big Bang nucleosynthesis, electro-weak precision measurements and accelerator searches. We show that the velocity broadening of the resonance plays an essential role for the possibility of the detection of a spectral feature originating from resonant photon-DM scattering. Depending upon the particle setup and the DM surface mass density, the favored range of DM particle masses lies between tens of keV and a few MeV, while the resonant photon absorption energy is predicted to be between tens of keV and few GeV. 

\end{abstract}




\maketitle

\section{Introduction}

One of the crucial properties of the {\em dark} matter (DM) is the feebleness of its coupling to the electromagnetic field. The early decoupling of DM from the baryon-photon fluid is a basic ingredient of the current picture of structure formation, and various direct DM detection experiments set stringent limits on the coupling of DM with ordinary matter. The phenomenological possibilities of a charged \cite{gould}, or a milli-charged \cite{milliconst,milli} DM species, or that of DM featuring an electric or magnetic dipole moment \cite{kris} were considered in several recent studies, all pointing towards a severe suppression of any coupling of the DM with photons. Significant absorption or scattering of photons by DM appears to ruled out, perhaps implying that {\em DM does not cast shadows}.

In this analysis we investigate the possibility that, while the typical scattering cross section of DM with photons is very small, photons with the right energy can resonantly scatter off DM particles.  We show that this resonant scattering might result in peculiar absorption features in the spectrum of distant sources. This effect can occur if the extension of the standard model of particle physics required to accommodate a (neutral) DM particle candidate $\chi_1$ also encompasses (1) a second, heavier neutral particle $\chi_2$ and (2) an electric and/or magnetic transition dipole moment which couples the electromagnetic field to $\chi_1$ and $\chi_2$. We also assume, for definiteness, that $\chi_1$ and $\chi_2$ are fermionic fields. In this setting, there exists a special photon energy $E_\gamma^{\rm res}$ where the scattering cross-section of photons by DM is resonantly enhanced to the unitarity limit. If the resonance is broad enough, and the cross section and DM column number density are large enough, the spectrum of distant photon sources might in principle feature a series of absorption lines corresponding to DM halos at different redshifts. If such anomalous absorption features exist, not only would they provide a smoking gun for the particle nature of DM, but they could also potentially give information about the distribution of DM in the Universe.

\section{The Model}

We adopt here a completely model-independent setting, where we indicate with $m_{1,2}$ the masses of $\chi_{1,2}$, and consider the effective interaction Lagrangian 
\begin{equation}\label{eq:lagr}
\mathcal L_{\rm eff}=-\frac{i}{2}\ \bar \chi_2\ \sigma_{\mu\nu}\frac{a+b\gamma_5}{\widetilde{M}}\chi_1\ F^{\mu\nu}.
\end{equation}
In the rest frame of $\chi_1$, the photon-DM scattering mediated through an $s$-channel $\chi_2$ exchange [see Fig.~\ref{fig:feyn}(a)] is resonant at the photon energy
\begin{equation}\label{eq:eres}
E^{\rm res}_{\gamma}=\frac{m_{2}^2-m_{1}^2}{2m_{1}}.
\end{equation}
For $E_\gamma\approx E^{\rm res}_{\gamma}$, the $\gamma$-DM scattering cross section can be approximated with the relativistic Breit-Wigner (BW) formula
\begin{equation}\label{eq:xsec}
\sigma_{\gamma\chi_1}(E_\gamma)=\frac{2\pi}{p_{\rm CM}^2}\  \frac{\Gamma_{\chi_2\rightarrow\chi_1\gamma}}{\Gamma_{\chi_2}}  \ \frac{(m_{2}\ \Gamma_{\chi_2})^2}{(s-m_{2}^2)^2+(m_{2}\ \Gamma_{\chi_2})^2},
\end{equation}
where $p_{\rm CM}$ indicates the modulus of the center-of-mass momentum, $s$ is the center-of-mass energy squared, $\Gamma_{\chi_2}$ is the total decay width of $\chi_2$, and $\Gamma_{\chi_2\rightarrow\chi_1\gamma}$ is the decay width of $\chi_2$ into $\chi_1$ and a photon. The value of $\sigma_{\gamma\chi_1}$ at $E_\gamma=E^{\rm res}_{\gamma}$ saturates the unitarity limit provided $B_{\chi_1\gamma}\equiv\Gamma_{\chi_2\rightarrow\chi_1\gamma}/\Gamma_{\chi_2}\approx1$. Under this assumption, even if $\chi_1$ and $\chi_2$ featured interactions different in their detailed microscopic nature from those described by Eq.~(\ref{eq:lagr}) (such as a transition milli-charge, or fermion-sfermion loops in neutralino DM models), the maximal resonant $\gamma$-DM scattering cross section would always be given by $\sigma_{\gamma\chi_1}(E_\gamma)$ for $E_\gamma\approx E^{\rm res}_\gamma$. From Eq.~(\ref{eq:lagr}), we compute
\begin{equation}
\Gamma_{\chi_2\rightarrow\chi_1\gamma}=\frac{|a|^2+|b|^2}{\pi {\widetilde{M}}^2}\left(\frac{m_{2}^2-m_{1}^2}{m_{2}}\right)^3.
\end{equation}
In the remainder of this study, for conciseness, we shall denote
\begin{equation}
M^2\equiv\frac{\pi {\widetilde{M}}^2}{|a|^2+|b|^2},
\end{equation}
and, in order to maximize the scattering rate of photons by DM, we will assume a model with $B_{\chi_1\gamma}\approx1$. All the quantities above can be trivially rephrased in terms of $m_2$ and of the two ratios $R\equiv m_1/m_2$ and $\eta\equiv m_2/M$ as
\begin{eqnarray}\label{eq:array}
&&\Gamma\equiv\Gamma_{\chi_2} = m_{2}\ \eta^2\ \left(1-R^2\right)^3\\[0.3cm]
&\nonumber &E^{\rm res}_{\gamma} =  m_{2}\ \frac{1-R^2}{2R} \quad \sigma_{\gamma\chi_1}(E^{\rm res}_{\gamma}) = \frac{8\pi}{\left(1-R^2\right)^2}\frac{1}{m_{2}^2}\\[0.3cm]
&&\nonumber \sigma_{\gamma\chi_1}(\tilde E\equiv \frac{E_\gamma}{m_{2}})=\frac{2\pi}{m_{2}^2}\frac{R+2\tilde E}{R\tilde E^2}\frac{\tilde\Gamma^2}{\left(R^2+2R\tilde E-1\right)^2+\tilde\Gamma^2}
\end{eqnarray}
where $\tilde\Gamma\equiv\Gamma/m_{2}$.

Let us now turn to the effects of the resonant scattering of photons emitted by a distant source. The mean specific intensity at the observed frequency $\nu_0$ as seen by an observer at redshift $z_0$ from the direction $\psi$ is given by
\begin{equation}
J(\nu_0,z_0,\psi)=\frac{(1+z_0)^3}{4\pi}\int_{z_0}^\infty{\rm d}z\left[\frac{{\rm d}l}{{\rm d}z}(z)\right]\epsilon(\nu,z,\psi)e^{-\tau_{\rm eff}},
\end{equation}
where
\begin{equation}
\nu=\nu_0\frac{1+z}{1+z_0},\qquad \frac{{\rm d}l}{{\rm d}z}(z)=\frac{c}{H(z)(1+z)},
\end{equation}
$\epsilon$ is the emissivity per unit comoving volume, and $\tau_{\rm eff}$ is the effective opacity. The latter can be cast as
\begin{equation}\label{eq:opacity}
\tau_{\rm eff}(\nu_0, z_0, z, \psi)=\int_{z_0}^z{\rm d}z^\prime\ \sigma(\nu^\prime)\frac{\rho_{}(z^\prime)}{m_{1}} \left[\frac{{\rm d}l}{{\rm d}z}(z^\prime)\right] \, ,
\end{equation}
where $\rho$ is the DM density.
To get a numerical feeling of whether the resonant scattering of photons leads to a sizable effect, we define
\begin{equation}\label{eq:tau}
\tau\equiv\frac{\sigma\ \Sigma_{\sss\rm DM}}{m_1}
\end{equation}
where $\Sigma_{\sss\rm DM}$ indicates an effective DM {\em surface density}, associated with the integral along the line of sight of the DM density. When the quantity $\tau\sim1$ in Eq.~(\ref{eq:tau}) for $E_\gamma=E^{\rm res}_\gamma$ we expect a significant absorption for photon energies $E_\gamma\approx E^{\rm res}_\gamma$.

Once a photon from a background source scatters off an intervening DM particle, the flux from the source itself is attenuated as long as the photon is diffused into an angle larger than the angular resolution of the instrument. The kinematics of the process closely resembles that of the relativistic Compton scattering \cite{kn} or Thompson scattering at lower energies. Roughly speaking, the relevant quantity can be cast as the fraction $F$ of scattered photons which end up being  scattered into an angle smaller than the instrumental angular resolution $\theta_{\rm\sss AR}$, over the total number of scattered photons. For an order of magnitude estimate it is easy to show that, apart from the details of the DM distribution geometry, $F$ depends on the two variables $\theta_{\rm\sss AR}$ and $x\equiv E_\gamma/m_{\chi_1}$. Making simple assumptions, we estimate the values of $F$ for an instrument featuring an angular resolution of one degree, over the range $10^{-2}<x<10^2$, fall within $10^{-4}\lesssim F\lesssim 5\times 10^{-3}$. We therefore can safely assume that if a photon scatters off DM, it is effectively lost ({\em i.e.} the flux of photons from the background source is effectively depleted by photon-DM scattering processes).

\section{Constraints on the parameter space}\label{sec:const}
\begin{figure}
\centerline{\psfig{file=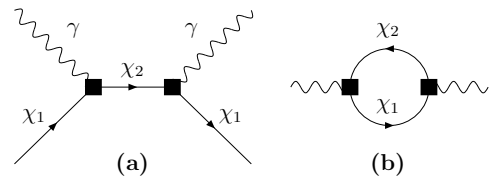,width=2.50in,angle=0}}
\caption{{\bf (a)}: Feynman diagram associated to the resonant photon scattering cross section described by Eq.~(\protect{\ref{eq:xsec}}); the black square indicates the effective vertex of the lagrangian (\protect{\ref{eq:lagr}}). {\bf (b)}: The $\chi_1-\chi_2$ loop diagram contributing to the vacuum polarization tensor (\protect{\ref{eq:vacu}}).}
\label{fig:feyn}
\end{figure}

Since the lagrangian (\ref{eq:lagr}) effectively couples the electromagnetic field to $\chi_1$ and $\chi_2$, depending upon the size of the coupling and the mass of the two particles $\chi_{1,2}$, the constraints that apply to a milli-charged particle $\psi$ ({\em e.g.} neutrinos featuring a small electric charge $e^\prime\ll e$ \cite{raffeltbook}, or the paratons of Ref.~\cite{milliconst}) will also be relevant for the present setting. 

\begin{figure}
\centerline{\psfig{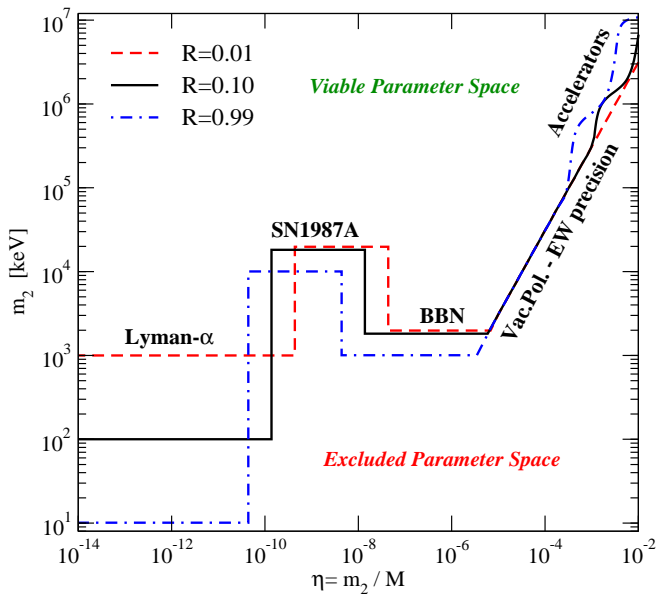}}
\caption{Constraints from Lyman-$\alpha$, SN 1987 A, BBN, Electroweak precision and other collider data on the model under consideration here, on the ($\eta,m_2$) plane, at $R=0.01,0.1$ and 0.99. The parameter space regions ruled out lie below the curves in the figure.}
\label{fig:fig2}
\end{figure}

The parameter space of the model we consider here consists of the parameters $m_2$, $\eta$ and $R$; for future convenience, we choose to represent the viable range of parameters on the ($\eta,m_2$) plane, at fixed representative values of $R=0.01,0.1$ and 0.99 [Fig.~\ref{fig:fig2}].

To translate the constraints from the milli-charged scenario in the present language, we need to compute the cross section $\sigma(f^+f^-\to\chi_1\chi_2)$, and compare it to the standard $\sigma(f^+f^-\to\overline\psi\psi)=\epsilon^2\sigma(f^+f^-\to e^+e^-)$ cross section, where $\epsilon\equiv e^\prime/e$. We find
\begin{eqnarray}
\sigma(f^+f^-\to\chi_1\chi_2)&=&\frac{16\alpha}{s}\ R\ \eta^2\times\\
&&\nonumber\sqrt{1+(1-R^2)^2\frac{m_2^4}{s^2}-2(1+R^2)\frac{m_2^2}{s}}.
\end{eqnarray}

A first simple astrophysical constraint on the model is based on avoiding excessive energy losses in stars that can produce $\chi_1\chi_2$ pairs by various reactions, in particular through plasma decay processes. The most stringent limits come from avoiding an unacceptable delay of helium ignition in low-mass red giants. The relevant energy scale in the process is the plasma frequency $\omega_P\approx8.6$ keV, and the limit applies, roughly, to masses $m_2\lesssim 2\omega_P/(1+R)$, constraining \cite{raffeltbook}
\begin{equation}
8\times 10^{-31}\lesssim R\eta^2\lesssim2\times 10^{-19}\quad{\rm [Red\ Giants]}. 
\end{equation}
While the lower limit stems from the energy losses argument, the upper limit comes from the requirement that the mean free path of $\chi_1$ is smaller than the physical size of the stellar core: if the $\chi_1$ particles get trapped, the impact on the stellar evolution through energy transfer would in any case be negligible compared to other mechanisms \cite{raffeltbook,milliconst}.

At such low masses, however, constraints from large scale structure, and namely from Lyman-$\alpha$ forest data, on the smallest possible mass for the DM particle force $Rm_2\gtrsim 10$ keV \cite{lya}. This bound corresponds to the left-most horizontal lines in Fig.~\ref{fig:fig2}.

For a narrow range of effective $\gamma-\chi_1-\chi_2$ couplings, the cooling limit discussed above can be applied to the SN 1987A data. For a SN core plasma frequency $\omega_P\approx 10$ MeV, values of $m_2\lesssim 2\omega_P/(1+R)$ can be ruled out in the range of couplings \cite{raffeltbook}
\begin{equation}
2\times 10^{-21}\lesssim R\eta^2\lesssim2\times 10^{-17}\quad{\rm [SN\ 1987A]}. 
\end{equation}
The limits from SN 1987A show up on the ($\eta,m_2$) plane as the rectangular regions of parameter space shown in Fig.~\ref{fig:fig2}.

If $\chi_{1,2}$ reach thermal equilibrium in the Early Universe before big bang nucleosynthesis (BBN), they contribute to the energy density and thus to the expansion rate. Translating in the present language the constraints from BBN found in Ref.~\cite{milliconst}, if $m_2\lesssim 2\times(1\ {\rm MeV})/(1+R)$ then 
\begin{equation}
R\eta^2\lesssim1.7\times 10^{-20}\quad{\rm [BBN]}. 
\end{equation}
The BBN limit corresponds to the central horizontal lines shown in Fig.~\ref{fig:fig2}.

As pointed out in \cite{kris}, a strong constraint on the size of DM magnetic or electric dipole moments is related to the size of the photon transverse vacuum-polarization tensor [see Fig.~\ref{fig:feyn}, (b)],
\begin{equation}\label{eq:vacu}
\Pi^{\mu\nu}(k^2)=\Pi(k^2)(k^2 g^{\mu\nu}-q^\mu q^\nu).
\end{equation}
The strongest constraint derived from (\ref{eq:vacu}) comes from the effect of the running of the fine-structure constant, for momenta ranging up to the $Z^0$ mass, on the relationship between $m_W$, $m_Z$ and $G_F$. Using (\ref{eq:lagr}), we computed
\begin{equation}
\Delta r=\Pi(m_Z^2)-\Pi(0)-k^2\left(\frac{\partial\Pi(k^2)}{\partial k^2}(0)\right),
\end{equation}
finding
\begin{eqnarray}\label{eq:loop}
&&\nonumber\Delta r=\frac{m_1 m_2}{\pi^2 M^2}\int_0^1{\rm d}x\Big[\ln\left(1-\frac{x(x-1)m_Z^2}{(x-1)m_2^2-x m_1}\right)+\\
&& -\frac{x(x-1)m_Z^2}{m_1 m_2}\frac{m_1 m_2+(1-x)^2m_Z^2}{(x-1)m_2^2-xm_1^2-x(x-1)m_Z^2}\Big].
\end{eqnarray}
The theoretically computed standard model values and the experimental inputs yield a limit on extra contributions to the running of $\alpha$, namely $\Delta r<0.003$ at 95\% C.L.~\cite{kris}. With particle masses $m_1, m_2\ll m_Z$, Eq.~(\ref{eq:loop}) reduces to
\begin{equation}\label{eq:asymbound}
\Delta r\simeq\frac{m_Z^2}{3\pi^2 M^2},
\end{equation}
implying $M\gtrsim3.4 m_Z$ for consistency with electroweak precision observables. The limits from Eq.~(\ref{eq:loop}) rule out the region below the line labeled as ``Vac.Pol.-EW precision'' in Fig.~\ref{fig:fig2}.

Lastly, high energy accelerator experiments also put constraints on particles with an effective coupling to photons. Such particles could have been seen in free quark searches \cite{Yao:2006px}, at the anomalous single photon (ASP) detector at the SLAC storage ring PEP \cite{asp} (designed to look for events in the form $e^+e^-\to\gamma\ +$ weakly interacting particles) and in beam-dump experiments from vector-meson decays and direct Drell-Yan production \cite{Golowich:1986tj}. The combination of all accelerator constraints rules out the relatively massive and strongly coupled models lying below the upper-right curvy lines on the ($\eta,m_2$) plane shown, for three values of $R$, in Fig.~\ref{fig:fig2}. 

This completes our discussion of the constraint on the parameter space of the model under consideration here: the viable parameter space, for a given $R$, lies {\em above} the lines shown in Fig.~\ref{fig:fig2}, while the portions of parameter space that are ruled out correspond to the regions of the plot {\em below} the various constraint lines.

\section{Resonant DM-Photon Scattering}

\begin{figure}
\centerline{\psfig{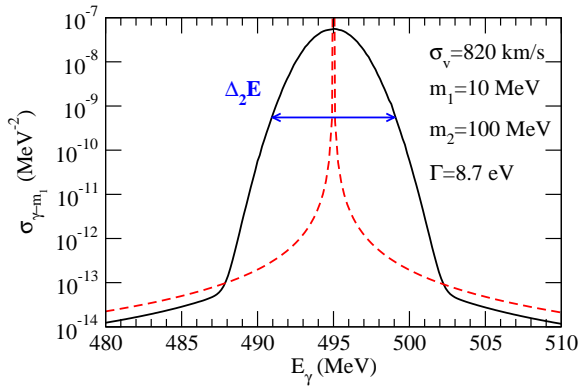}}
\caption{The photon-DM cross section at zero velocity (red dashed line) and with a velocity-dispersion $\sigma_v=820\,${\rm km/s} (black line), for $m_1=10$ MeV, $m_2=100$ MeV and $\Gamma=8.7$ eV.}
\label{fig:resonance}
\end{figure}
Since DM particles live in halos characterized by a velocity dispersion $\sigma_v$, which depending upon the mass of the DM halo can take values from roughly $100\, ${\rm km/s} to over $1000\, ${\rm km/s}, the momentum distribution of the  DM particles approximately follows a Maxwell-Boltzmann distribution,
\begin{equation}\label{eq:distribution}
f_T(p)=\sqrt{\frac{2}{\pi}}\frac{p^2{\rm e}^{-p^2/(2a^2)}}{a^3},\quad a=m_1 \sigma_v.
\end{equation}
An incoming photon will therefore scatter off DM particles with the above momentum distribution, and the ``{\em effective}'' scattering cross section will be given by the following average:
\begin{align}\label{eq:convolution}
\sigma^T_{\gamma\chi_1}(E_\gamma)=\int_0^\infty {{\rm d}p\ f_T(p)}\left<\sigma\right>_{\mu} \, ,
\end{align}
where
\begin{align}\label{eq:anginte}
\left<\sigma\right>_{\mu} = \int_{-1}^{1}\frac{d\mu}{2}\frac{2 \pi }{p_{\rm CM}^2}\frac{(m_{2}\Gamma_{\chi_2})^2}{\left(s-m_{2}^2\right)^2+(m_{2}\Gamma_{\chi_2})^2},
\end{align}
where $\mu$ is the cosine of the incident DM-$\gamma$ angle, and where the center of mass energy and momentum squared read
\begin{align}
s=m_{1}^2 + 2 E_{\gamma}\left(\sqrt{p^2+m_{1}^2}-\mu p \right)
\end{align}
and
\begin{align}
p_{\rm CM}^{-2} = \frac{4 s}{(m_1^2-s)^2}.
\end{align}
The integral in Eq.~(\ref{eq:anginte}) can be solved analytically, and we report the result in the Appendix. As a result of the averaging procedure of Eq.~(\ref{eq:convolution}), the maximum of the effective cross section is no longer the peak value $\sigma(E^{\rm res}_{\gamma})$, but will be a non-trivial combination of the latter, $\Gamma_{\chi_2}$, $\sigma_v$, and $E_{\gamma}^{\rm res}$. We illustrate an instance of the result of the broadening of the BW cross section in Fig.~\ref{fig:resonance}.

\begin{figure}
\centerline{\psfig{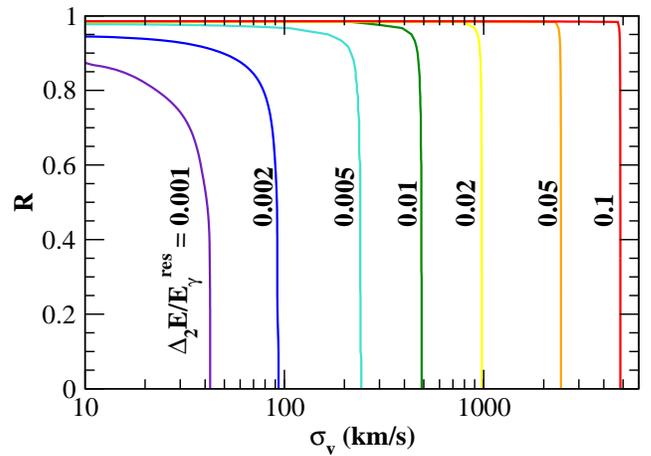}}
\caption{Curves of constant $\Delta_2 E/E_\gamma^{\rm res}$ ({\em i.e.} the relative width of the DM-photon resonant scattering cross section) on the $(\sigma_v,R)$ plane.}
\label{fig:sigmat}
\end{figure}
Given an instrument with an energy resolution $\xi$, defined as the relative energy resolution ({\em i.e.} the ratio of the energy resolution at the energy $E$ over the energy $E$ itself), we require the width $\Delta_2 E$ of the resonance (which we define, for convenience, to be the range of values of $E_\gamma$ where $\sigma_{\gamma\chi_1}(E_\gamma)>10^{-2}\sigma(E^{\rm res}_\gamma)$) to be at least as large as $\xi\times E_\gamma^{\rm res}$. 

To a good approximation, the solution to the equation $\Delta_2 E=\xi\times E_\gamma^{\rm res}$ is independent of $\eta$, since $\sigma^T_{\gamma\chi_1}(E_\gamma)\propto\Gamma_{\chi_2}$ for $E_\gamma\approx E_\gamma^{\rm res}$. Also, since $\Delta_2 E\propto m_1$ [see Eq.~(\ref{eq:distribution})], at fixed $R$ and small $\Gamma_{\chi_2}$, the ratio $\Delta_2 E/E_\gamma^{\rm res}$ is independent of $m_2$ as well. We therefore plot, in Fig.~\ref{fig:sigmat}, curves at constant values of $\Delta_2 E/E_\gamma^{\rm res}$ on the $(\sigma_v,R)$ plane. As clear from Eq.~(\ref{eq:distribution}), the larger the value of the velocity dispersion $\sigma_v$, the larger $\Delta_2 E$. From Eq.~(\ref{eq:array}) we also understand that, as $R\to1$, $E_\gamma^{\rm res}\to0$, explaining why arbitrarily large values of $\Delta_2 E/E_\gamma^{\rm res}$ can be obtained for large $R$ [see the upper part of Fig.~\ref{fig:sigmat}].

\begin{figure}
\centerline{\psfig{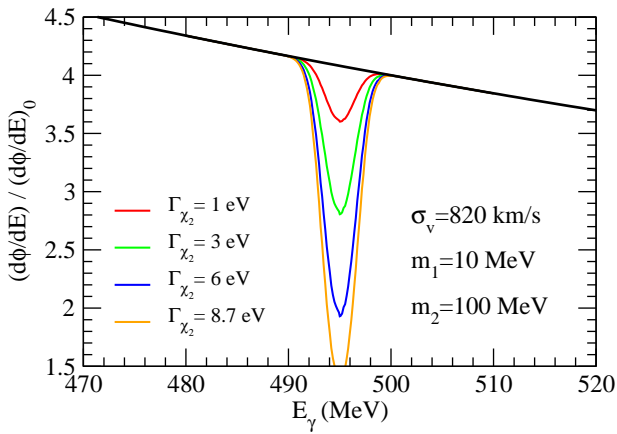}}
\caption{The effect of photon-DM scattering on a putative power-law spectrum (see Eq.~(\protect{\ref{eq:spectru}})) for a source located behind (or at the center of) a cluster with features comparable to those of the Coma cluster. We assume $m_{\chi_1}=10$ MeV, $m_{\chi_2}=100$ MeV and four different values of the ``{\em resonance width}'' $\Gamma_{\chi_2}$.}
\label{fig:flux}
\end{figure}
How would the spectrum of a background source look like after photons have resonantly scattered off DM? We address this question in Fig.~\ref{fig:flux}. We assume for definiteness (our results don't critically depend upon the particular spectral shape) a power-law spectrum of the form
\begin{equation}\label{eq:spectru}
\frac{{\rm d}\phi}{{\rm d}E}=\left(\frac{{\rm d}\phi}{{\rm d}E}\right)_0\times\left(\frac{E}{1\ {\rm GeV}}\right)^{-2}.
\end{equation}
We consider a setup where $m_{\chi_1}=10$ MeV and $m_{\chi_2}=100$ MeV, and as an example we focus on the case of a source located behind (or at the center of) a cluster with features similar to those of the Coma cluster. Making use of the estimates provided in Ref.~\cite{Colafrancesco:2005ji}, we consider a DM surface density (integrating the D05 DM profile \cite{d05} along the direction of the center of the cluster, within one virial radius of the cluster center) of $\Sigma_{\sss\rm DM}\simeq5\times10^{29}\ {\rm MeV}/{\rm cm}^2$. Also, we assume a velocity dispersion of $\sigma_v=820\,${\rm km/s}. Notice that the redshift of the Coma cluster, $z\simeq0.0231$, is small enough that the effect of photon redshift on the shape and location of the absorption feature is completely negligible. Making use of these estimates, the effect on the background source spectrum depends entirely upon the value of $\Gamma_{\chi_2}$: for large values of the latter quantity the DM halo is opaque to photons with energies around $E_\gamma^{\rm res}$.

We show in Fig.~\ref{fig:flux} how the spectrum defined in Eq.~(\ref{eq:spectru}) is affected by setups with various different values of $\Gamma_{\chi_2}$. For $\Gamma_{\chi_2}\gtrsim6$ eV (orange line), the absorption is almost complete around $E_\gamma^{\rm res}$. Smaller values of  $\Gamma_{\chi_2}$ imply only a partial deformation of the spectrum, and a reduced energy range where absorption effectively takes place. For $\Gamma_{\chi_2}\lesssim 1$ eV the absorption feature would be almost invisible.

\begin{figure}
\centerline{\psfig{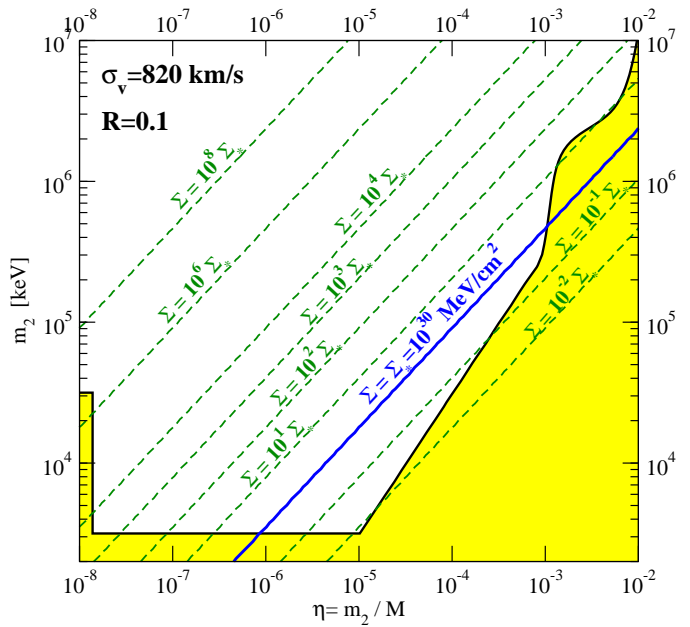}}
\caption{The ($\eta,m_2$) plane at $R=0.1$ and $\sigma_v=820\,${\rm km/s}. The region shaded in yellow is ruled out by the constraints discussed in Sec.~\ref{sec:const}. The green dashed lines correspond to fixed values of the DM surface density required to have a substantial absorption feature in the spectrum of distant sources, in units of $\Sigma_*\equiv10^{30}\ {\rm MeV}/{\rm cm}^2$ (blue solid line).}
\label{fig:fig6}
\end{figure}
We summarize our results on the $(\eta,m_2)$ plane in Fig.~\ref{fig:fig6}, for the same reference values we employed in Fig.~\ref{fig:sigmat}, {\em i.e.} $R=0.1$ and $\sigma_v=820\,${\rm km/s}. For this choice of parameters, $\Delta_2 E/E_\gamma^{\rm res}\approx0.015$. The area shaded in yellow at the bottom right of the plot is ruled out by the various constraints discussed in the previous section. The green dashed lines correspond to fixed values for $\Sigma_{\sss\rm DM}$ such that $\tau\sim1$ in Eq.~(\ref{eq:tau}), in units of $\Sigma_*\equiv 10^{30}\ {\rm MeV}/{\rm cm}^2$ (blue solid line). For DM surface densities $\Sigma_{\sss\rm DM}\lesssim{\cal O}(10^{30}\ {\rm MeV}/{\rm cm}^2)$, absorption is possible for DM particle masses $m_1=Rm_2\approx{\cal O}(0.3$--$50)$ MeV. The absorption feature, in this plot, is predicted according to Eq.~(\ref{eq:eres}) to occur at a photon energy $E_\gamma^{\rm res}\simeq 5m_2$; henceforth, in the range above, we predict $E_\gamma^{\rm res}\approx{\cal O}(15$--$2500)$ MeV.

The analogue of Fig.~\ref{fig:fig6} for other values of $\sigma_v$ and $R$ can be directly read out of our results shown in Fig.~\ref{fig:sigmat} taking into account the constraints shown in Fig.~\ref{fig:fig2}, and the fact that the values of $\Sigma_{\sss\rm DM}$ such that $\tau\sim1$ in Eq.~(\ref{eq:tau}) scale approximately linearly with $R$. For instance, again for DM surface densities $\Sigma_{\sss\rm DM}\lesssim 10^{30}\ {\rm MeV}/{\rm cm}^2$, we would predict a DM particle mass $20\ {\rm keV}\lesssim m_1\lesssim 50$ MeV for $R=0.01$, and around 1--50 MeV for $R=0.99$. Analogously, the location of the absorption feature is predicted in the range $E_\gamma^{\rm res}\approx$ 1 MeV to 150 GeV for $R=0.01$, and at $E_\gamma^{\rm res}\approx$ 10--500 keV  for $R=0.99$. In the present setup, therefore, for reasonable DM surface densities, the location of the absorption feature varies in a wide range of photon energies, from tens of keV up to several GeV.

\section{Discussion}

Photons from background sources will in general pass through various DM halos at all intermediate redshifts, resulting in a cumulative cosmological effect leading, in principle, to a broadening and modulation of the absorption feature described above. In Ref.~\cite{Ullio:2002pj}, for instance, an analogous computation was carried out for the monochromatic photon emission from DM pair annihilations into two photons; the detailed setup here is however different as the effect depends linearly rather than quadratically on the dark matter density distribution. A similar cosmological broadening was also discussed for the case of resonant $\nu\bar\nu\rightarrow Z^0$ high energy neutrino absorption {\em e.g.} in Ref.~\cite{Barenboim:2004di}. The spatial homogeneity of the cosmic neutrino background results however in a completely different column density structure than in the present case. The detailed computation of this cumulative cosmological effect depends on several assumptions about the distribution and nature of DM structures in the Universe, on the presence of DM clumps or other substructures~\cite{Profumo:2006bv}, and on the assumed halo density profiles and velocity distributions~\cite{Ullio:2002pj}.  We leave the detailed analysis of this effect to a future study.

In passing we note that thermally-produced DM candidates with masses in the tens of keV to the MeV range, often referred to as {\em warm} DM candidates, exhibit potentially interesting features in structure formation, suppressing, through free-streaming, small scale structures, and partly alleviating the cusp problem of cold DM models (see {\em e.g}. Ref.~\cite{warmdm} and references therein).  Depending on the details of the particle physics model constraints on such warm DM candidates might be used to constrain our scenario.

Closing the photon line in Fig.~\ref{fig:feyn} {\bf (a)} into a loop generates radiative corrections to $m_1$. If the latter are too large, the values we employed must be corrected accordingly, and small values of $m_1$ might not be theoretically allowed. We can estimate the size of these corrections as
\begin{equation}
\frac{\delta m_{1}}{m_{1}}\approx\frac{1}{16\pi^2}\frac{m_{1}^3}{m_{2}\ M^2}=\frac{R^3\eta^2}{16\pi^2}.
\end{equation}
Radiative corrections are therefore smaller than $10^{-2}$ provided $\eta\lesssim 1$, a condition which is always widely satisfied in the parameter space under consideration here. A mass mixing term would also be generated by the interaction responsible for the effective lagrangian (\ref{eq:lagr}); in principle, one should then rotate Eq.~(\ref{eq:lagr}) to the proper mass-eigenstate basis. However, the relative size of the induced $\chi_1-\chi_2$ mixing is also very suppressed, as it roughly scales as $R\eta^2$, and can be thus safely neglected here.

In the scenario we are discussing here, the $\chi_1$ particles can also pair annihilate into two photons through a $\chi_2$ $t$- or $u$- channel exchange. The resulting cross section can be estimated as
\begin{equation}
\sigma_{\gamma\gamma}\approx\left(\frac{m^2_{1}}{M^2\ m_{2}}\right)^2=\frac{R^4\eta^4}{m_{2}^2}
\end{equation}
Pair annihilation of $\chi_1$'s into photons can \emph{a priori} be the process through which DM annihilates in the early Universe and potentially this could thermally produce the amount of DM inferred in the current cosmological standard model. In the range of couplings and masses we obtain here, the above mentioned annihilation channel is insufficient to produce a large enough pair annihilation rate in the Early Universe in order to get the required DM abundance $\Omega_{\chi_1}\approx\Omega_{\rm CDM}$. Other channels, otherwise irrelevant for the present discussion, and compatible with the present setting, can however contribute to give the $\chi_1$ particles the right pair annihilation rate

The same diagram discussed above, and the same pair annihilation cross section, intervene in the pair annihilation rate of $\chi_1$'s today into monochromatic photons of energy $E_{\gamma\gamma}=m_{\chi_1}=Rm_{2}$. The flux of photons per unit solid angle from monochromatic pair annihilations of $\chi_1$'s can be written as
\begin{equation}
\phi_\gamma(\theta,\Delta\Omega)\approx\frac{R^2\eta^4}{m_2^2}\frac{J(\theta,\Delta\Omega)}{4\pi}
\end{equation}
where, in this instance, the quantity $J$ refers to the following line-of-sight integral along the direction $\theta$ averaged over the solid angle $\Delta\Omega$
\begin{equation}
J(\theta,\Delta\Omega)\equiv\frac{1}{\Delta\Omega}\int_{\Delta\Omega}{\rm d}\Omega\int_{\rm l.o.s.}{\rm d}l\ \rho^2(l).
\end{equation}
When $m_1<m_2\ll m_Z$, we can derive an upper limit to the monochromatic photon flux which is independent of $m_{1,2}$, namely
\begin{equation}
\phi_\gamma(\theta,\Delta\Omega)\lesssim0.3R^2\left(\frac{J(\theta,\Delta\Omega)}{{\rm GeV}^4\ {\rm cm}^{-3}}\right)\ {{\rm cm}^{-2}\ {\rm s}^{-1} {\rm sr}^{-1}}
\end{equation}
Taking an angular region $\Delta\Omega=10^{-3}$ sr in the direction of the galactic center the range of values which $J$ can take for various viable DM halo models is $10^{-5}$--$10^{-2}\ {\rm GeV}^4\ {\rm cm}^{-3}$. This means that one expects a flux of monochromatic photons in the range $(10^{-6}$--$10^{-3})R^2\ {{\rm cm}^{-2}\ {\rm s}^{-1} {\rm sr}^{-1}}$. The diffuse gamma-ray flux in the Galactic center region as measured by COMPTEL and EGRET \cite{comptelegret} is at the level of 0.01 ${{\rm cm}^{-2}\ {\rm s}^{-1} {\rm sr}^{-1}}$ at a gamma-ray energy of 1-3 MeV. Extrapolating to smaller energies we expect an even larger flux at energies around or smaller than 100 keV.  This makes it extremely hard to reconstruct a would be annihilation signal from the galactic background. Dedicated searches for line emissions show that instruments such as INTEGRAL-SPI also fail to achieve the sensitivity required here \cite{Teegarden:2006ni}. On the other hand, this also means that the class of models discussed above is not currently constrained by monochromatic photon emissions. Furthermore, observations of objects where the diffuse gamma-ray background is expected to be suppressed, such as the nearby dwarf galaxies \cite{draco}, can potentially lead to constraints or even to the detection of the monochromatic emission line predicted here.

If, as we describe here, photons scatter off DM at significant rates, one might also expect other associated features besides the absorption lines and the monochromatic emissions described above. Scattering off DM might generate an effective ``{\em index of refraction}'' in the photon propagation, possibly inducing {\em e.g.} time delays in transient sources at different frequencies, or frequency-dependent distortions of the photon paths for steady sources. A detailed discussion of these effects lies, however, beyond the scopes of the present analysis.

Neutralino DM in the context of the minimal supersymmetric extension of the Standard Model (MSSM) can in principle produce an effective lagrangian setup as that in Eq.~(\ref{eq:lagr}), for instance through fermion-sfermion loops coupling two different neutralinos $\tilde\chi_1$ and $\tilde\chi_i$, $i>1$. From the discussion above, however, it is clear that supersymmetric DM cannot produce any sizable photon absorption. First, the lightest supersymmetric particle (LSP) in any viable low energy supersymmetry setup is typically heavier than at least a few GeV (for exceptions, {\em e.g.} in the next-to-MSSM, see \cite{nmssm}). This implies, as can be read off Fig.~\ref{fig:fig6}, very large values of $\Sigma_{\sss\rm DM}$ to get $\tau\sim1$ in Eq.~(\ref{eq:tau}). Secondly, the assumptions we made at the beginning that $B_{\chi_1\gamma}\approx1$ does not hold in general in the MSSM: the radiative $\tilde\chi_2\rightarrow\tilde\chi_1\gamma$ decay can be the dominant mode only in restricted regions of parameter space, {\em e.g.} when phase space suppresses other three-body decay modes. The resulting effective $\eta$, in the notation set above, is in any case limited from above by 
\begin{equation}
\eta\approx\frac{eg^2}{16\pi^2}\lesssim 10^{-3}
\end{equation}
Requiring $\Gamma_{\tilde\chi_2}\simeq\Gamma_{\tilde\chi_2\rightarrow\tilde\chi_1\gamma}$ implies $m_{\tilde\chi_2}\simeq m_{\tilde\chi_1}$, and typically $R\gtrsim0.995$. Furthermore, since in the MSSM when two neutralinos are quasi degenerate the lightest chargino is also quasi degenerate with them, LEP2 limits on the chargino mass \cite{Yao:2006px} force $m_{\tilde\chi_1}\gtrsim 100$ GeV. $R\gtrsim0.995$ also implies $\Gamma_{\tilde\chi_2}\lesssim10^{-10}$ GeV. These values for the model parameters imply ({\em a}) small relative widths and ({\em b}) too large DM surface densities for the absorption feature to be detectable. Relaxing the requirement that $B_{\chi_1\gamma}\approx1$ would not help anyway, since the cross section (\ref{eq:xsec}) receives the large suppression factor $\Gamma_{\tilde\chi_2\rightarrow\tilde\chi_1\gamma}/\Gamma_{\tilde\chi_2}$, and the photon absorption process is again suppressed.

One can envision, however, various particle physics scenarios where the phenomenology described above can take place. For instance, a concrete particle physics setup which can explain at once the DM abundance, neutrino masses and mixing, the baryon asymmetry of the Universe and, potentially, inflation, is the so-called $\nu$MSM \cite{nmsm}, or one of its extensions \cite{nmsmext}. These models feature a light quasi-stable sterile neutrino with a mass in the tens of keV \cite{nmsm} up to $\mathcal O$(10) MeV \cite{nmsmext} range, and heavier Majorana neutrinos with a mass at the GeV scale. Extending this class of models with an effective interaction of the form of our Eq.~(\ref{eq:lagr}) gives rise to the phenomenology described above and, hence, to possible resonant photon scattering. 

\section{Conclusions}

We have shown that photons can, in principle, resonantly scatter off DM, through an effective lagrangian featuring a dipole transition moment coupling photons, the DM particle $\chi_1$ and a heavier neutral particle $\chi_2$. We discussed the constraints on the model from stellar energy losses, data from SN 1987A, the Lyman-$\alpha$ forest, Big Bang nucleosynthesis, electro-weak precision measurements and accelerator searches. The effective resonant ``absorption'' cross section is broadened by the effect of the momentum distribution of DM particles in DM halos. We showed that DM particles in the tens of keV to a few MeV range can lead to resonant photon scattering (resulting in absorption lines which can lie between tens of keV up to tens of GeV) provided the DM surface mass density is at least of ${\cal O}(10^{28}\ {\rm MeV}/{\rm cm}^2)$. We also pointed out that typical supersymmetric DM (the weak-scale neutralino) does not cast any shadows ({\em i.e.} it does not ``absorb'' photons), while photon absorption can take place in other particle physics setups which can explain various pieces of physics beyond the standard model. 

\begin{acknowledgments}
We thank John Beacom and Christopher Hirata for insightful comments on an earlier draft of this manuscript.  We thank Vincenzo Cirigliano, Shane Davis, Mikhail Gorshteyn, Tesla Jeltema, Marc Kamionkowski and Enrico Ramirez-Ruiz for related discussions. SP is supported in part by DoE grants DE-FG03-92-ER40701 and DE-FG02-05ER41361, and NASA grant NNG05GF69G. KS is supported by NASA through Hubble Fellowship grant HST-HF-01191.01-A awarded by the Space Telescope Science Institute, which is operated by the Association of Universities for Research in Astronomy, Inc., for NASA, under contract NAS 5-26555. 
\end{acknowledgments}

\appendix*

\section{The Angular Integral $\left<\sigma\right>_{\mu}$}
The angular integral $\left<\sigma\right>_{\mu}$ can be computed analytically with the result
\begin{align}
\left<\sigma\right>_{\mu}&=& \frac{\pi (m_{2}\Gamma_{\chi_2})^2}{E_{\gamma}^2\left[(\Delta m^2)^2+m_{2}^2\Gamma_{\chi_2}^2\right]}\times\\[0.3cm]
&&\nonumber\left\{ 2 + \frac{E_{\gamma}}{p}\left[c_f(f_1-f_2)+c_g g\right]\right\} \,
\end{align}
where
\begin{align}
\Delta m^2 = m_{2}^2 - m_{1}^2 \,
\end{align}
\begin{align}
c_f = 1 + \frac{2 m_{1}^2 \Delta m^2}{(\Delta m^2)^2 + m_{2}^2 \Gamma_{\chi_2}^2} \,
\end{align}
\begin{align}
c_g &=& \frac{2\Delta m^2}{m_{2}\Gamma_{\chi_2}}\left[1+\frac{\Delta m^2 (m_2^2+m_1^2)}{(\Delta m^2)^2 + m_{2}^2 \Gamma_{\chi_2}^2}\right]\\[0.3cm]
&&\nonumber-\frac{ (2m_{1}^2 - \Delta m^2)m_{2}\Gamma_{\chi_2}}{(\Delta m^2)^2 + m_{2}^2\Gamma_{\chi_2}^2} \,
\end{align}
\begin{align}
f_1 = \ln\left[\frac{(2 E_{\gamma}\sqrt{p^2+m_{1}^2}+ 2 E_{\gamma}p)^2}{(2 E_{\gamma}\sqrt{p^2+m_{1}^2}-2 E_{\gamma} p)^2}\right] \, ,
\end{align}
\begin{align}
f_2 = \ln\left[\frac{(2 E_{\gamma}\sqrt{p^2+m_{1}^2}+ 2 E_{\gamma}p- \Delta m^2)^2+m_{2}^2\Gamma_{\chi_2}^2}{(2 E_{\gamma}\sqrt{p^2+m_{1}^2}-2 E_{\gamma} p - \Delta m^2)^2+m_{2}^2\Gamma_{\chi_2}^2}\right] \, ,
\end{align}
and
\begin{align}
g&=&\arctan\left[\frac{\Delta m^2+2E_\gamma \left(p-\sqrt{m_1^2+p^2}\right)}{m_2\Gamma_{\chi_2}}\right]\\[0.3cm]
&&\nonumber - \arctan\left[\frac{\Delta m^2-2E_\gamma \left(p+\sqrt{m_1^2+p^2}\right)}{m_2\Gamma_{\chi_2}}\right].
\end{align}

\end{document}